\documentclass[
    final            
  4apaper]
  {aipproc}

\layoutstyle{8x11single}

    \usepackage{bm}

    \usepackage{amsmath}

     \usepackage{amssymb}

\newcommand\beq{\begin{equation}}
\newcommand\eeq{\end{equation}}
\newcommand\beqa{\begin{eqnarray}}
\newcommand\eeqa{\end{eqnarray}}
\newcommand{\nn}{\nonumber\\}

\newcommand{\ma}{m_i}
\newcommand{\mb}{m_j}
\newcommand{\cca}{\mathbf{v}_i}

\newcommand{\ccb}{\mathbf{v}_j}

\newcommand{\kk}{\widehat{\bm{\sigma}}}
\newcommand{\wwa}{\bm{\omega}_i}
\newcommand{\wwb}{\bm{\omega}_j}
\newcommand{\wa}{{\omega}_i}

\newcommand{\ww}{\bm{\omega}}
\newcommand{\Ia}{I_i}

\newcommand{\da}{\sigma_i}
\newcommand{\db}{\sigma_j}
\newcommand{\ds}{\sigma}
\newcommand{\dab}{\sigma_{ij}}

\newcommand{\gh}{\mathbf{v}_{ij}}

\newcommand{\een}{\alpha_{ij}}
\newcommand{\esn}{\alpha}
\newcommand{\eet}{\beta_{ij}}
\newcommand{\est}{\beta}

\newcommand{\qab}{\kappa_{ij}}
\newcommand{\qa}{\kappa_{i}}
\newcommand{\qb}{\kappa_{j}}
\newcommand{\q}{\kappa}
\newcommand{\fa}{f_{i}}

\newcommand{\far}{f_{i}^\text{rot}}
\newcommand{\fbr}{f_{j}^\text{rot}}
\newcommand{\Tat}{T_{i}^\text{tr}}
\newcommand{\Tbt}{T_{j}^\text{tr}}
\newcommand{\Tt}{T^\text{tr}}
\newcommand{\Tar}{T_{i}^\text{rot}}
\newcommand{\Tbr}{T_{j}^\text{rot}}
\newcommand{\Tr}{T^\text{rot}}

\newcommand{\Qab}{J_{ij}}
\newcommand{\Iab}{\mathcal{J}_{ij}}

\newcommand{\na}{n_i}
\newcommand{\nb}{n_j}

\newcommand{\zabt}{\xi_{ij}^\text{tr}}
\newcommand{\zt}{\xi^\text{tr}}
\newcommand{\zabr}{\xi_{ij}^\text{rot}}
\newcommand{\zr}{\xi^\text{rot}}
\newcommand{\al}{i}
\newcommand{\be}{j}
\newcommand{\tr}{\text{tr}}
\newcommand{\rot}{\text{rot}}
\newcommand{\fab}{f_{ij}}
\newcommand{\chiab}{{\chi}_{ij}}

\begin{document}

\title{Homogeneous Free Cooling State in Binary Granular Fluids of Inelastic Rough Hard Spheres}


\author{Andr\'es Santos}{
  address={Departamento de F\'{\i}sica, Universidad de
Extremadura, E-06071 Badajoz, Spain} }

\begin{abstract}
In a recent paper [A. Santos, G. M. Kremer, and V. Garzó, \emph{Prog. Theor. Phys. Suppl.} \textbf{184}, 31-48 (2010)] the collisional energy production rates associated with the translational  and rotational granular temperatures in a granular fluid mixture of inelastic rough hard spheres have been derived. In the present paper the energy production rates are explicitly decomposed into equipartition rates (tending to make all the temperatures equal) plus genuine cooling rates (reflecting the collisional dissipation of energy). Next the homogeneous free cooling state of a binary mixture is analyzed, with special emphasis on the quasi-smooth limit. A previously reported singular behavior (according to which a vanishingly small amount of roughness has a finite effect, with respect to the  perfectly smooth case, on the asymptotic long-time translational/translational temperature ratio) is further elaborated. Moreover, the study of the time evolution of the temperature ratios shows that
this dramatic influence of roughness already appears in the transient regime for times comparable to the relaxation time of perfectly smooth spheres.

\end{abstract}

\maketitle


\section{Introduction}
The simplest model of a granular fluid consists of a system of identical, inelastic smooth hard spheres with a constant coefficient of normal restitution \cite{G03}. Obviously, the model can be made closer to reality by introducing extra ingredients. In particular, polydispersity and roughness are especially relevant because they unveil an inherent breakdown of energy equipartition in granular fluids, even in homogeneous and isotropic states \cite{GD99,SD07,BT02,LHMZ98,GNB05,Z06}.

	In this work a fluid mixture of inelastic rough hard spheres characterized by  mutual coefficients of normal restitution $\{\alpha_{ij}\}$ and tangential restitution $\{\beta_{ij}\}$ is considered. First, a recent derivation by kinetic-theory arguments \cite{SKG10} of the collisional energy production rates ($\zabt$  and $\zabr$) associated with the translational and rotational temperatures ($\Tat$  and $\Tar$) is recalled and the energy production rates are decomposed into equipartition and cooling rates. Next, the results are applied to a binary mixture and the asymptotic long-time values and the time evolution of the three independent temperature ratios $\Tt_2/\Tt_1$, $\Tr_2/\Tr_1$, and $\Tr_1/\Tt_1$ and are analyzed in the homogeneous and isotropic free cooling state Special attention is paid to a paradoxical effect: the asymptotic long-time values of the temperature ratio $\Tt_2/\Tt_1$ in the nearly-smooth limit ($\beta_{ij}\to-1$)  differ from those obtained for purely smooth spheres ($\beta_{ij}=-1$).

\section{Collisional energy production rates\label{sec2}}
Let us consider a multi-component granular gas made of hard spheres of masses $\{\ma\}$, diameters $\{\da\}$, and moments of inertia $\{\Ia=\frac{1}{4}\ma\da^2\qa\}$.
The value of the dimensionless parameter $\qa$ depends on the mass distribution within the sphere and runs from the extreme values $\qa=0$ (mass concentrated on the center) to $\qa=\frac{2}{3}$ (mass concentrated on the surface).
Collisions between a sphere of component $i$ and a sphere of component $j$ are characterized by a coefficient of \emph{normal} restitution $\alpha_{ij}$ and a coefficient of \emph{tangential} restitution $\beta_{ij}$.
The former coefficient ranges from $\een=0$ (perfectly inelastic particles) to $\een=1$ (perfectly elastic particles), while the latter runs from $\eet=-1$ (perfectly smooth particles) to $\eet=1$ (perfectly rough particles). The total  (translational plus rotational) kinetic energy is dissipated upon collisions unless $\een=1$ and $\eet=\pm 1$.
The particular case of one-component systems has been widely studied in the literature (see, for instance, Refs.\ \cite{LHMZ98,GNB05,Z06}).

Let $\fa(\mathbf{r}_i,\cca,\wwa;t)$ and $\fab(\mathbf{r}_i,\cca,\wwa;\mathbf{r}_j,\ccb,\wwb;t)$ be the one-body and two-body distribution functions, respectively, where $\cca$ is the velocity of the center of mass and $\wwa$ is the angular velocity.
By starting from the Liouville equation and following standard steps, one can derive the Bogoliubov--Born--Green--Kirkwood--Yvon (BBGKY) hierarchy. The first equation of the hierarchy reads \cite{SKG10}
\beq
\partial_t \fa(\cca,\wwa)+\cca\cdot\nabla \fa (\cca,\wwa)=\sum_\be \Qab[\cca,\wwa|\fab],
\label{2}
\eeq
\beq
\Qab[\cca,\wwa|\fab]=\dab^2\int d\ccb\int d\wwb\int d\kk\, \Theta(\gh\cdot\kk)(\gh\cdot\kk)\left[\frac{1}{\een^2\eet^2}\fab(\mathbf{r}_i,\cca'',\wwa'';\mathbf{r}_i^-,\ccb'',\wwb'')-
\fab(\mathbf{r}_i,\cca,\wwa;\mathbf{r}_i^+,\ccb,\wwb)\right].
\label{III.2}
\eeq
Here, $\Qab$ is the collision operator, $\dab\equiv (\da+\db)/2$, $\gh\equiv\cca-\ccb$, $\mathbf{r}_i^\pm\equiv\mathbf{r}_i\pm\dab\kk$, and the double primes denote pre-collisional quantities giving rise to unprimed quantities as post-collisional values. The explicit form of the {restituting} collision rules can be found in Ref.\ \cite{SKG10}.

Given an arbitrary one-body  function $\psi_i(\cca,\wwa)$, we define its average as
\beq
\langle \psi_i(\cca,\wwa)\rangle\equiv \frac{1}{\na}\int d\cca \int d\wwa\, \psi_i(\cca,\wwa) \fa(\cca,\wwa), \quad \na=\int d\cca \int d\wwa\,  \fa(\cca,\wwa),
\label{III.3}
\eeq
where $\na$ is the number density of component $\al$. The \emph{rate of change} of the quantity $\psi_i(\cca,\wwa)$ due to collisions with particles of component $\be$ is
\beq
\Iab[\psi_i(\cca,\wwa)]\equiv\frac{1}{\na}\int d\cca\int d\wwa\, \psi_i(\cca,\wwa) \Qab[\cca,\wwa|\fab].
\label{3}
\eeq
In particular, the \emph{partial} temperatures associated with the translational and rotational degrees of freedom are
\beq
\Tat=\frac{\ma}{3}\langle (\cca-\mathbf{u})^2\rangle,\quad \Tar=\frac{\Ia}{3}\langle \wa^2\rangle,
\label{III.11}
\eeq
where
$\mathbf{u}={\sum_\al \ma\na\langle \cca\rangle}/{\sum_\al \ma\na}$
is the flow velocity. The corresponding collisional rates of change define the (partial) collisional \emph{energy production rates} as
\beq
\xi_i^\tr=\sum_\be \zabt,\quad
\xi_i^\rot=\sum_\be \zabr,\quad\zabt\equiv -\frac{\ma}{3\Tat}\Iab[(\cca-\mathbf{u})^2],\quad \zabr\equiv -\frac{\Ia}{3\Tar}\Iab[\wa^2].
\label{54}
\eeq
The \emph{global} temperature of the gas and its associated net \emph{cooling} rate are
\beq
T=\sum_\al\frac{\na}{2n}\left(\Tat+\Tar\right), \quad \zeta=\sum_{\al,\be} \frac{\na}{2nT}
\left(\Tat\zabt+\Tar\zabr\right),
\label{III.13}
\eeq
where $n=\sum_\al \na$ is the total number density. Note that, as usually done  in the literature on granular gases, the Boltzmann constant has been absorbed in the definitions of $\Tat$, $\Tar$, and $T$, so that these quantities have dimensions of energy.

\begin{figure}
  \includegraphics[height=.1\textheight]{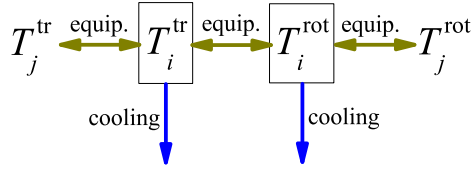}
\caption{Scheme on the two classes of contributions (equipartition rates and cooling rates) to the energy production rates $\zabt$ and $\zabr$ characterizing the effect on $\Tat$ and $\Tar$, respectively, of collisions with particles of component $j$ \cite{L09}.
\label{fig1}}
\end{figure}

The energy production  rates $\zabt$ and $\zabr$ defined in Eq.\ \eqref{54} do not have a definite sign. They can be decomposed into two classes of terms: equipartition rates and cooling rates (see Fig.\ \ref{fig1}). The equipartition terms, which exist even when energy is conserved by collisions ($\een=1$ and $\eet=\pm 1$), tend to make temperatures equal \cite{UKAZ09}. Therefore, they can be positive or negative depending on the signs of the differences $\Tat-\Tbt$, $\Tar-\Tbr$, and $\Tat-\Tar$. On the other hand, the cooling terms reflect the collisional energy dissipation and thus they are positive if $\een<1$ and/or $|\eet|<1$, vanishing otherwise. Only the cooling terms in $\zabt$ and $\zabr$ contribute to the net cooling rate  $\zeta$.

The quantities $\Iab[\psi_i(\cca,\wwa)]$  are functionals of the two-body distribution function $\fab$ and therefore they are generally rather intricate.
Now, let us imagine that, instead of the full knowledge of $\fab$, we only have at our disposal the local values of the two densities ($\na$ and $\nb$) and the four partial temperatures ($\Tat$, $\Tar$, $\Tbt$, and $\Tbr$). Then, we can get reasonable \emph{estimates} of $\Iab[\psi_i(\cca,\wwa)]$ from the replacement
\beq
\fab(\cca,\wwa;\ccb,\wwb)\to\na\nb\chiab \left(\frac{\ma\mb}{4\pi^2\Tat\Tbt}\right)^{3/2}\exp\left[-\ma\frac{(\cca-\mathbf{u})^2}{2\Tat}-\mb\frac{(\ccb-\mathbf{u})^2}{2\Tbt}\right]
\far(\wwa)\fbr(\wwb),
\label{IV.1}
\eeq
where $\chiab$ is the contact value of the pair correlation function. Equation \eqref{IV.1} can be justified by maximum-entropy arguments, except that here  a Maxwellian form for the rotational probability densities $\far(\wwa)$ and $\fbr(\wwb)$ is not needed. The expressions for $\zabt$ and $\zabr$ derived from the replacement \eqref{IV.1}  and the assumption $\langle \cca\rangle=\langle \ccb\rangle$ can be found in Ref.\ \cite{SKG10}. Here, those expressions are recast into sums of equipartition and cooling rates as follows:
\beq
\zabt=\frac{1+\een}{2}\xi_{\al\be}^{(1)}+\frac{1+\eet}{2}\left[\xi_{\al\be}^{(2)}+\qa\frac{\Tar}{\Tat}\xi_{\al\be}^{(3)}\right]+
\zeta_{\al\be}^{\tr,1}+\qa\frac{\Tar}{\Tat}\zeta_{\al\be}^{\rot},\quad \zabr=\frac{1+\eet}{2}\xi_{\al\be}^{(3)}+\zeta_{\al\be}^{\rot},
\label{100}
\eeq
\beq
\xi_{\al\be}^{(1)}=\frac{10}{3}\nu_{\al\be}\frac{\ma\mb}{(\ma+\mb)^2}\frac{\Tat-\Tbt}{\Tat},\quad \xi_{\al\be}^{(2)}=\frac{10}{3}\nu_{\al\be}\frac{\mb}{\ma+\mb}\frac{\qab}{1+\qab}\frac{\Tat-\Tar}{\Tat},
\label{102}
\eeq
\beq
\xi_{\al\be}^{(3)}=\frac{10}{3}\frac{\nu_{\al\be}}{\Tar}\frac{\ma\mb}{\qa\qb(\ma+\mb)^2}\left(\frac{\qab}{1+\qab}\right)^2
\left[\Tar-\Tbr+\qb\left(\Tat-\Tbt\right)+\qb(1+\mb/\ma)\left(\Tar-\Tat\right)\right],
\label{105}
\eeq
\beq
\zeta_{\al\be}^{\tr,1}=\frac{5}{6}\nu_{\al\be}\frac{1-\een^2}{\Tat}\frac{\ma\mb^2}{(\ma+\mb)^2}\left(\frac{\Tat}{\ma}+\frac{\Tbt}{\mb}\right),
\label{104}
\eeq
\beq
\zeta_{\al\be}^{\rot}=\frac{5}{6}\nu_{\al\be}\frac{1-\eet^2}{\Tar}\frac{\ma\mb^2}{\qa(\ma+\mb)^2}\left(\frac{\qab}{1+\qab}\right)^2
\left(\frac{\Tat}{\ma}+\frac{\Tbt}{\mb}+\frac{\Tar}{\ma\qa}+\frac{\Tbr}{\mb\qb}\right).
\label{106}
\eeq
In these equations, $\qab\equiv \qa\qb(\ma+\mb)/(\qa\ma+\qb\mb)$,
\beq
\nu_{\al\be}\equiv\frac{8\sqrt{2\pi}}{5}\chiab\nb{\dab^2}\sqrt{\frac{\Tat}{\ma}+\frac{\Tbt}{\mb}}
\label{56b}
\eeq
is an effective collision frequency, and we have assumed $\langle \ww_i\rangle=\langle \ww_j\rangle=\mathbf{0}$.
The quantities $\xi_{\al\be}^{(1,2,3)}$ represent \emph{equipartition} rates. They do not have a definite sign and vanish if all the temperatures are equal. The equipartition rate $\xi_\tr^{(1)}$ is always present (even for perfectly elastic spheres, $\een=1$) and tends to equilibrate both translational temperatures. The rates $\xi_{\al\be}^{(2)}$ and $\xi_{\al\be}^{(3)}$ do not contribute in the case of smooth spheres ($\eet=-1$). The former tends to equilibrate the translational and rotational temperatures of component $\al$, while the latter tends to equilibrate both rotational temperatures but is also affected by the other temperature differences. The quantities $\zeta_{\al\be}^{\tr,1}$ and $\zeta_{\al\be}^{\rot}$, on the other hand, are positive definite and represent \emph{cooling} rates. The former   only vanishes if the spheres are elastic, whilst the latter only vanishes if the spheres are  either perfectly smooth ($\eet=-1$) or perfectly rough ($\eet=1$). It is straightforward to check that ${\na}\Tat\xi_{\al\be}^{(1)}+{\nb}\Tbt\xi_{\be\al}^{(1)}=0$ and $\na\left[\Tat\xi_{\al\be}^{(2)}+(1+\qa)\Tar\xi_{\al\be}^{(3)}\right]
+\nb\left[\Tbt\xi_{\be\al}^{(2)}+(1+\qb)\Tbr\xi_{\be\al}^{(3)}\right]=0$.
Therefore, as expected, the equipartition rates do not contribute to the net cooling rate. The latter is given by
\beqa
\zeta&=&\sum_{\al\be} \frac{\na}{2nT}
\left[\Tat\zeta_{\al\be}^{\tr,1}+(1+\qa)\Tar\zeta_{\al\be}^{\rot}\right]\nn
&=&\frac{5}{24}\sum_{\al\be} \frac{\na}{nT}\nu_{\al\be}\frac{\ma\mb}{\ma+\mb}\left[(1-\een^2)\left(\frac{\Tat}{\ma}+\frac{\Tbt}{\mb}\right)+\frac{\qab}{1+\qab}
(1-\eet^2)
\left(\frac{\Tat}{\ma}+\frac{\Tbt}{\mb}+\frac{\Tar}{\ma\qa}+\frac{\Tbr}{\mb\qb}\right)\right].
\label{114}
\eeqa

\section{Homogeneous free cooling state}
In the  homogeneous and isotropic free cooling state  the evolution equations for the total and partial temperatures are
\beq
 \partial_t T=-\zeta T,\quad \partial_t\ln\frac{\Tat}{T}=-\left(\zt_\al-\zeta\right),\quad
\partial_t\ln\frac{\Tar}{T}=-\left(\zr_\al-\zeta\right).
\label{57}
\eeq
The first equation describes the monotonic decrease of temperature (unless $\een=1$ and $\eet=\pm 1$ for all the pairs). The remaining equations show that the temperature ratios $\Tat/T$ and $\Tar/T$ decrease or increase, depending on the sign of the differences $\zt_\al-\zeta$ and $\zr_\al-\zeta$, respectively. After a certain transient stage, those temperature ratios eventually reach constant values, while the total temperature keeps decreasing following Haff's law $T(t)=T(t_0)\left[1+\zeta(t_0)t/2\right]^{-2}$. In this asymptotic regime one has equal production rates, i.e., $\zt_1=\zt_2=\cdots=\zt_s=\zr_1=\zr_2=\cdots=\zr_s$, where $s$ is the number of components. Use of  expressions \eqref{100}--\eqref{106} allows one to get an algebraic set of equations whose solution gives the asymptotic long-time values of the temperature ratios.

In the particular case of a binary mixture ($s=2$), it is convenient to choose the three relevant temperature ratios as $\Tt_2/\Tt_1$, $\Tr_2/\Tr_1$, and $\Tr_1/\Tt_1$. The parameter space is twelve-dimensional: the three coefficients of normal restitution ($\esn_{11}$, $\esn_{12}$, $\esn_{22}$), the three coefficients of tangential restitution ($\est_{11}$, $\est_{12}$, $\est_{22}$), the two moment-of-inertia parameters ($\q_1$ and $\q_2$), the mass ratio ($m_2/m_1$), the size ratio ($\ds_2/\ds_1$), the concentration ratio ($n_2/n_1$) and the total packing fraction $\phi=(\pi/6)(n_1\ds_1^3+n_2\ds_2^3)$.

For the sake of illustration, here we  particularize to an equimolar mixture where all the spheres are uniformly solid  and are made of the same material, the size of the spheres of one component being twice that of  the other component. More specifically, $n_1=n_2$, $\esn_{11}=\esn_{12}=\esn_{22}=\esn$, $\est_{11}=\est_{12}=\est_{22}=\est$, $\q_1=\q_2=\frac{2}{5}$, $\ds_2/\ds_1=2$, and $m_2/m_1=8$. Moreover, we consider a dilute granular gas ($\phi\ll 1$), so that $\chi_{\al\be}\approx 1$. Thus only the parameters $\esn$ and $\est$ remain free.

In Ref.\ \cite{SKG10} it was observed that the asymptotic value of the translational/translational temperature ratio $\Tt_2/\Tt_1$ exhibits a peculiar behavior in the smooth-sphere limit  $\est\to -1$: it tends to a finite value different from (in fact higher than) the value directly obtained in the case of perfectly smooth spheres \cite{GD99}.
Thus, a tiny amount of roughness has  dramatic effects on the temperature ratio $\Tt_2/\Tt_1$, producing an enhancement of non-equipartition.
This singular behavior of the case $\beta=-1$ in the free cooling state is further elaborated in this paper.

\begin{figure}
  \includegraphics[width=\columnwidth]{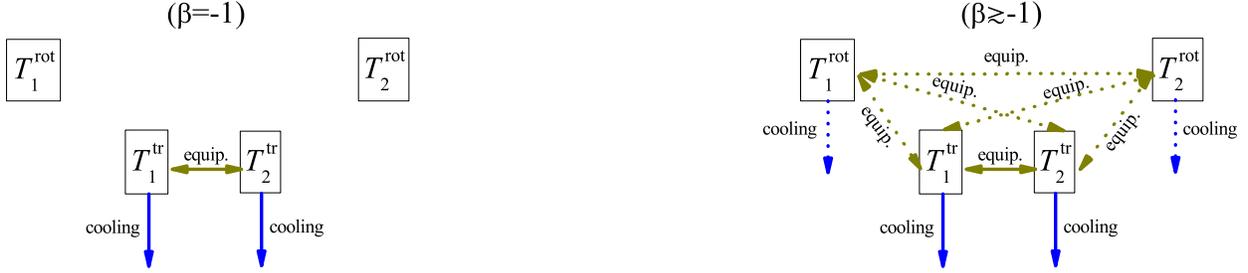}
\caption{Schematic representation of the energy transfer channels in the free cooling state of a binary mixture.
The left diagram corresponds to perfectly smooth particles ($\beta=-1$), in which case both rotational temperatures are completely decoupled each other and also from the translational temperatures, so they do not change with time.  The right diagram corresponds to quasi-smooth particles ($\beta\gtrsim -1$), in which case weak channels of energy transfer exist and also the rotational temperatures are subject to a weak cooling.
\label{fig2}}
\end{figure}

The physical origin of the singularity is the following one. If one strictly has $\beta=-1$, the rotational degrees of freedom are ``frozen'', so that the rotational temperatures are finite while the translational temperatures keep decreasing. The rotational/translational temperature ratios diverge but, since the spheres are perfectly smooth, there is no mechanism transferring energy from the rotational to the translational degrees of freedom. This situation is represented by the left diagram of Fig.\ \ref{fig2}. On the other hand, if $\beta=-1+\epsilon$, where $0<\epsilon\ll 1$, the huge rotational/translational temperature ratios are eventually able to ``feed'' the  weak energy channels connecting the rotational and translational temperatures (see the right diagram of Fig.\ \ref{fig2}), thus producing a non-negligible effect (``ghost'' effect) on the ratio $\Tt_2/\Tt_1$.

Let us see this in more detail. In the limit $\est\to -1$ one can see that the rotational/translational temperature ratios diverge as
${\Tr_1}/{\Tt_1}\approx \theta_1 \epsilon^{-1}$ and ${\Tr_2}/{\Tt_1}\approx \theta_2 \epsilon^{-2}$.
It can be found that the energy production rates behave as $\xi_1^\tr/\nu_{11}\approx F_1^\tr(\Tt_2/\Tt_1,\theta_2)$, $\xi_2^\tr/\nu_{11}\approx F_2^\tr(\Tt_2/\Tt_1,\theta_2)$, $\xi_1^\rot/\nu_{11}\approx F_1^\rot(\Tt_2/\Tt_1,\theta_2/\theta_1)\epsilon$, and $\xi_2^\rot/\nu_{11}\approx F_2^\rot(\Tt_2/\Tt_1)\epsilon$, where $F_{1,2}^{\tr,\rot}$ are explicit functions of the indicated arguments. Since both rotational production rates vanish in the limit $\epsilon\to 0$, the condition of equal production rates implies that $F_1^\tr(\Tt_2/\Tt_1,\theta_2)=0$,
$F_2^\tr(\Tt_2/\Tt_1,\theta_2)=0$, and $F_1^\rot(\Tt_2/\Tt_1,\theta_2/\theta_1)=F_2^\rot(\Tt_2/\Tt_1$). The first two conditions yield $\Tt_2/\Tt_1$ and $\theta_2$. Once obtained, the third condition gives $\theta_1$.
In the pure smooth-sphere problem, only the translational degrees of freedom matter and therefore the temperature ratio $\Tt_2/\Tt_1$ is  obtained from the condition $\xi^\tr_{1}=\xi^\tr_{2}$ with $1+\beta=0$, which becomes $F_1^\tr(\Tt_2/\Tt_1,0)=F_2^\tr(\Tt_2/\Tt_1,0)$. Obviously, the obtained solution differs from that given by $F_1^\tr(\Tt_2/\Tt_1,\theta_2)=F_2^\tr(\Tt_2/\Tt_1,\theta_2)=0$.

In the preceding discussion we have implicitly assumed that the spheres are inelastic ($\esn<1$). In the perfectly elastic case ($\esn=1$),  the scalings for the temperature ratios are quite different and turn out to be $\Tt_2/\Tt_1=1+\gamma\epsilon$, $\Tr_1/\Tt_1=\vartheta_1 \epsilon$, and $\Tr_2/\Tt_1=\vartheta_2 \epsilon$. The production rates behave as $\xi_1^\tr/\nu_{11}\approx \Phi_1^\tr(\gamma)\epsilon$, $\xi_2^\tr/\nu_{11}\approx \Phi_2^\tr(\gamma)\epsilon$, $\xi_1^\rot/\nu_{11}\approx \Phi_1^\rot(\vartheta_1)\epsilon$, and $\xi_2^\rot/\nu_{11}\approx \Phi_2^\rot(\vartheta_2)\epsilon$, where $\Phi_{1.2}^{\tr,\rot}$ are linear functions. The coefficient $\gamma$ is simply obtained from $\Phi_1^\tr(\gamma)=\Phi_2^\tr(\gamma)$. Once $\gamma$ is known, $\vartheta_1$  and $\vartheta_2$  follow from $\Phi_1^\rot(\vartheta_1)=\Phi_1^\tr(\gamma)$ and $\Phi_2^\rot(\vartheta_1)=\Phi_2^\tr(\gamma)$, respectively.

\begin{figure}
  \includegraphics[width=\columnwidth]{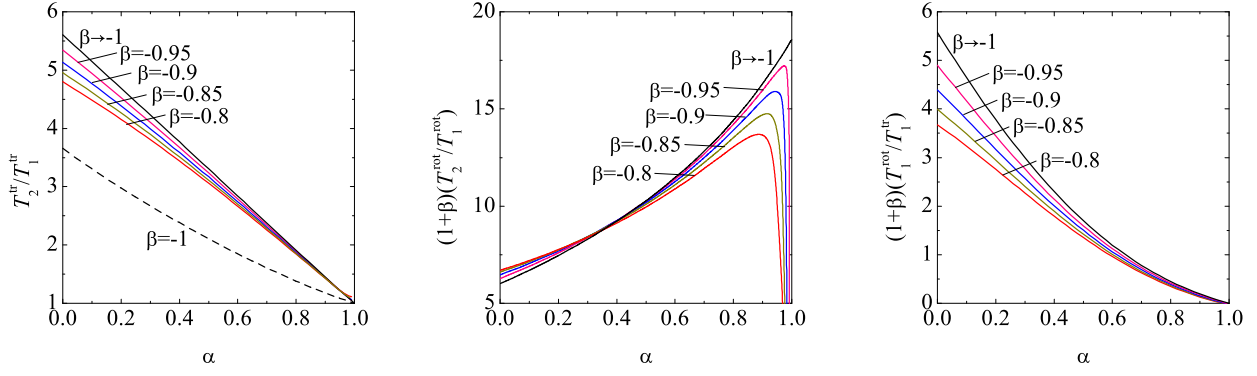}
\caption{Plot of the asymptotic long-time values of the temperature ratios $\Tt_2/\Tt_1$ (left panel), $(1+\beta)\Tr_2/\Tr_1$ (middle panel), and $(1+\beta)\Tr_1/\Tt_1$ (right panel) as functions of $\alpha$ for $\beta=-0.8$, $-0.85$, $-0.9$, $-0.95$, and in the limit $\beta\to -1$. In the left panel the translational/translational ratio for the strict case of perfectly smooth spheres ($\beta=-1$) is also plotted.
\label{fig3}}
\end{figure}
Figure \ref{fig3} shows the quantities $\Tt_2/\Tt_1$, $(1+\beta)\Tr_2/\Tr_1$, and $(1+\beta)\Tr_1/\Tt_1$ as functions of $\alpha$ for several values of $\beta$, including the limit $\beta\to -1$. It is clearly observed that the translational/translational temperature ratio in the limit $\beta\to -1$ differs from the values obtained in the perfectly smooth case \cite{GD99}. In the latter case, moreover, the rotational temperatures $\Tt_1$ and $\Tr_2$ maintain their initial values and thus the scaled ratio  $(1+\beta)\Tr_2/\Tr_1$ vanishes, while $(1+\beta)\Tr_1/\Tt_1$ is indetermined.

\begin{figure}
  \includegraphics[width=\columnwidth]{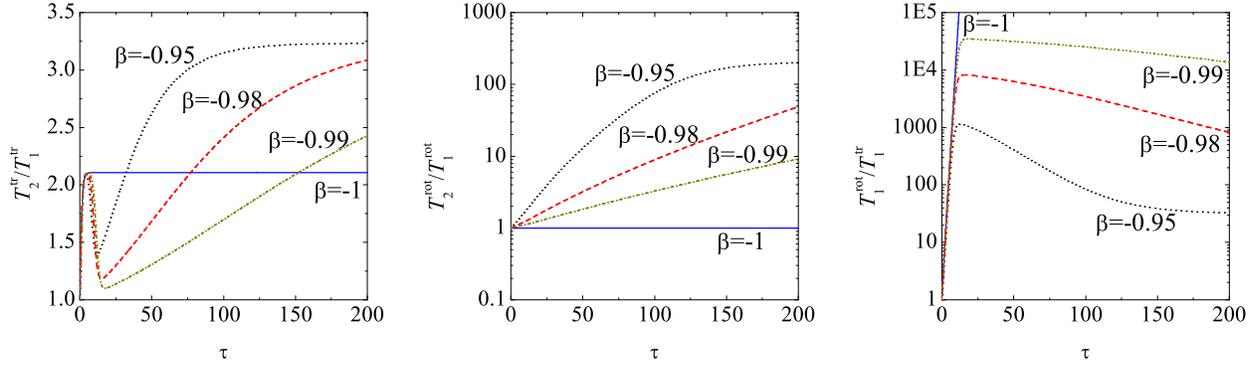}
\caption{Time evolution of the temperature ratios $\Tt_2/\Tt_1$ (left panel), $\Tr_2/\Tr_1$ (middle panel), and $\Tr_1/\Tt_1$ (right panel)  for $\alpha=0.5$ and $\beta=-0.95$, $-0.98$, $-0.99$,  and $-1$.
 The initial state is an equipartition one, i.e., $\Tt_1(0)=\Tt_2(0)=\Tr_1(0)=\Tr_2(0)$.
\label{fig4}}
\end{figure}
Since the scaled energy production rates $\lim_{t\to\infty}\xi_{1,2}^{\tr,\rot}/\nu_{11}$  go to zero in the quasi-smooth limit $\beta\to -1$, it is reasonable to wonder whether the paradoxical singular effect, $\lim_{t\to\infty}\left.\Tt_2/\Tt_1\right|_{\beta\to-1}\neq \lim_{t\to\infty}\left.\Tt_2/\Tt_1\right|_{\beta=-1}$, is apparent only for times much longer than the characteristic relaxation time of $\left.\Tt_2/\Tt_1\right|_{\beta=-1}$.
In order to investigate this possible scenario, the set of three coupled equations $\partial_\tau \ln(\Tt_2/\Tt_1)=-(\xi_2^\tr-\xi_1^\tr)/\nu_{11}$, $\partial_\tau \ln(\Tr_1/\Tt_1)=-(\xi_1^\rot-\xi_1^\tr)/\nu_{11}$, and $\partial_\tau \ln(\Tr_2/\Tt_1)=-(\xi_2^\rot-\xi_1^\tr)/\nu_{11}$ has been numerically solved for several values of $\beta >-1$, the solution being compared with that of $\partial_\tau \ln(\Tt_2/\Tt_1)=-(\xi_2^\tr-\xi_1^\tr)/\nu_{11}$ at $\beta=-1$. Here, $\tau=\int_0^t dt'\,\nu_{11}(t')$ is a measure of time in units of the accumulated number of 1--1 collisions. Figure \ref{fig4} shows the time evolution of the three temperature ratios for $\alpha=0.5$ and some representative values of $\beta$. We can observe that in the perfectly smooth case ($\beta=-1$) the translational/translational temperature ratio monotonically increases until reaching a plateau value at $\tau\approx 4$, the rotational/rotational ratio keeps its initial value, and the rotational/translational ratio increases exponentially with $\tau$ (Haff's law). The important feature in the cases $\beta\neq -1$ is that, while  the relaxation time needed to reach the asymptotic long-time values actually increases as $1+\beta$ decreases, the curves for $\Tt_2/\Tt_1$ and $\Tr_1/\Tt_1$ markedly differ from the case of perfectly smooth spheres just after $\tau\approx 4$. Therefore, the dramatic difference between the quasi-smooth limit ($\beta\to -1$) and the perfectly smooth case ($\beta=-1$) manifests itself not only in the asymptotic long-time regime but also in the transient regime  for times comparable to the relaxation time of perfectly smooth spheres.

\section{Conclusion}
In this paper the homogeneous free cooling state of a binary granular gas made of inelastic rough hard spheres has been analyzed. Special attention has been paid to the quasi-smooth limit ($\beta\to -1$), where an interesting singular phenomenon appears. It turns out that a vanishingly small amount of roughness has a significant effect on the translational/translational temperature ratio, with respect to the strict perfectly smooth case ($\beta=-1$).  This paradoxical (``ghost'') phenomenon is essentially due to the fact that, even if $\est$ is close to $-1$, the transfer of energy from the rotational to the translational degrees of freedom  eventually becomes activated when the rotational temperatures are much larger than the translational.

It is important to remark that this singular behavior of the smooth case is closely tied to the non-stationary character of the free cooling state and thus disappears in homogeneous \emph{steady} states. For instance, in a granular gas driven by a white-noise thermostat, the stationarity conditions are $({\Tt_1}/{m_1})\xi^\tr_{1}=({\Tt_2}/{m_2})\xi^\tr_{2}$ and $
 \xi^\rot_{1}=\xi^\rot_{2}=0$. It is straightforward to check that in this case the  ratio $T_2^\text{tr}/T_1^\text{tr}$ in the limit $\est\to -1$ coincides with that of $\est= -1$ \cite{BT02}.


\begin{theacknowledgments}
Insightful discussions with G. M. Kremer, V. Garz\'o, and J. W. Dufty are gratefully acknowledged.   This work has been supported by the
Ministerio de Educaci\'on y Ciencia (Spain) through Grant No.\
FIS2007--60977 (partially financed by FEDER funds) and by the Junta
de Extremadura (Spain) through Grant No.\ GR10158.
\end{theacknowledgments}



\bibliographystyle{aipproc}   

\end{document}